# Generating a Simulation-Based Contacts Matrix for Disease Transmission Modeling at Special Settings

Mahdi M. Najafabadi [a]*, Ali Asgary [a], Mohammadali Tofighi [a], and Ghassem Tofighi [b]

*Author Affiliations:* [a] *Advanced Disaster, Emergency and Rapid Response Simulation (ADERSIM), York University, Toronto, Canada;* [b] *School of Applied Computing, Sheridan College, Toronto, Canada*

## Abstract

*Since a significant amount of disease transmission occurs through human-to-human or social contact, understanding who interacts with whom in time and space is essential for disease transmission modeling, prediction, and assessment of prevention strategies in different environments and special settings. Thus, measuring contact mixing patterns, often in the form of a contacts matrix, has been a key component of heterogeneous disease transmission modeling research. Several data collection techniques estimate or calculate a contacts matrix at different geographical scales and population mixes based on surveys and sensors. This paper presents a methodology for generating a contacts matrix by using high fidelity simulations which mimic actual workflow and movements of individuals in time and space. Results of this study show that such simulations can be a feasible, flexible, and reasonable alternative method for estimating social contacts and generating contacts mixing matrices for various settings under different conditions.*

## 1. Introduction and Background

Person-to-person disease transmission is largely driven by who interacts with whom in time and space (Prem et al., 2020). Past and current studies, especially those conducted during the COVID-19 pandemic have further revealed that individuals who have more person-to-person contact per period have higher chances of becoming infected, infecting others, and being infected earlier during pandemics (Smieszek et al., 2014). Moreover, during pandemics, contact patterns can drastically shift from their baseline conditions due to individuals' actions and public health measures (Prem et al., 2020). Therefore, estimating the baseline contact patterns, as well as how they would shift over time, have been important components of disease transmission modelling and predictions (Hens et al., 2009). These patterns are often generated in the form of contact matrices and are used in mathematical models to develop 'Who Acquires Infection from Whom' (WAIFW) matrices which determine the force of infections from infected individuals to susceptible populations (Beutels et al., 2006). The accuracy of the disease models and their effectiveness in public health decisions during disease outbreaks and pandemics very much depends on access to, and availability of, detailed and readily available data about contact mixing patterns (McCarthy et al., 2020).

The more knowledge we have about the contact mixing patterns, the better we can understand and predict infectious disease dynamics and assess the effects of various public health interventions and measures that aim to control the spread of directly transmitted infections (Iozzi et al., 2010; Smieszek et al., 2014). Measuring contact mixing patterns has been an important, challenging, and expensive part of disease modeling research. During the past two decades, several methods have been proposed and used by researchers for calculating social contacts at different geographical scales and population mixes. These methods can be classified into survey-based, device-based, model-based, and simulation-based methods.

Surveys are among the most widely used techniques in contacts measurement research and practice (Smieszek et al., 2014). Surveys collect contact data among individuals at different age groups or socio-economic settings (home, school, workplace, community). Using surveys, participants are asked to record or recall all contacts that they have had with others in all spaces they have gone to or visited during a specified period of time, such as a day or week. Surveys collect data through direct observation, diaries, questionnaires, direct interviews, phone interviews, and web-based questionnaires. The POLYMOD study used a survey-based approach in eight European countries. Participants were asked to complete a contact diary to record details about all the people they met over the course of a single day (Eames et al., 2011; Grantz et al., 2020). The contacts, along with the demographic and geographic data collected from the participants, were then used to develop baseline contact patterns (Hens et al., 2009).

Device-based methods, on the other hand, collect contact data through a variety of wearable devices and sensors, usually radio-frequency identification (RFID) tags or smartphones. Some of these devices keep a log of the physical location of individuals per time-step (i.e., the interval by which they record the physical location data). This data is then analyzed to find proximities among different individuals. Other types of devices only record or transmit the data regarding the proximity of individuals with relevant timestamps which allow the calculation of duration for each recorded effective contact. The use of wearable devices is a practical way of calculating contacts in small settings such as schools, workplaces, and hospitals (Champredon et al., 2018; Duval et al., 2018). Moreover, advancements in smartphone technology and wider access to them have made large-scale device-based contact measurement faster, more accurate, and more feasible. Collecting digital device-based data is a quick alternative in situations where survey data is hard to collect or unavailable. Some recent studies have demonstrated the successful collection of device-based contact data at different scales (Watson et al., 2017). However, this type of contact measurement faces several technical, financial, privacy, and accuracy challenges that limit its utilization. More specifically, there is a large inequity of access to these devices and technologies around the world.

Model-based contact mixing studies often use data collected from surveys and sensors to customize and replicate social contact matrices for other communities and settings. They also try to mathematically model particular communities or settings to calculate potential contacts.

Many of the current COVID-19 related disease transmission models in different countries and regions use mathematical approaches to recreate contact matrices for desired communities and settings based on the POLYMOD study (McCarthy et al., 2020).

Finally, the simulation-based method aims to recreate detailed synthetic or use high-fidelity simulations of desired settings to estimate potential human contacts in them. An example of this approach is the *Little Italy* simulation that created a synthetic society to reconstruct contact data using an individual-based model (IBM) and Time Use Survey data (Iozzi et al., 2010). A comparison of contact matrices generated by this simulation with those developed for Italy by the POLYMOD study showed that simulation-based matrices can provide a fruitful complementary approach to survey-based matrices. With advances in simulation software and hardware, especially agent-based and discrete event simulations, and the availability of supporting data through other methods and tools, this approach is promising. Especially in situations where the accessing of special settings is not quite feasible due to privacy concerns or ongoing outbreaks. Despite their high potential, the use of simulation-based methods for this purpose has been very limited to date.

Given the increased computational power at lowered costs, simulation-based methods are becoming an even stronger alternative for conventional research methods. Moreover, these simulation models can have a user-interface component that allows policymakers to use them as a decision-support tool, especially in healthcare and pandemic control (Asgary, Najafabadi, et al., 2020; Asgary et al., 2021).

In this paper, we contribute to filling this gap by suggesting a novel hybrid simulation-based methodology. This methodology utilizes a combination of agent-based and discrete event simulations, to generate contact mixing matrices based on existing and predefined workflows, individual schedules, and behaviors in special settings. While we have already applied this method in some healthcare settings, the method introduced here can be easily replicated in other settings where human interaction takes place. This method can be applied in schools, hospitals, shopping malls, long-term care facilities, offices, manufacturing facilities, places of worship, sports facilities, museums, factories, etc., to generate contact matrices.

Generating a contact matrix from simulations is accomplished in four steps:

1. Develop a 2D/3D model of the setting
2. Define agents' workflow and behavior in the setting
3. Run the simulation and record fine-grained agent-to-agent data
4. Generate contacts mixing matrices

The rest of this paper explains each of these steps in detail in three examples:

1. A simulation model of a hospital hemodialysis ward that utilizes this method.
2. A simulated Intensive Care Unit (ICU) that uses a Bayesian version of compound probabilities in calculation of the matrix.
3. A simulation of a large gathering that allows examining this method for a much larger number of agents.

## 2. Developing the Simulation Model

### 2.1. Defining the Physical Layouts

In the first step, the floor plan of the selected setting is used to create a 2D and/or 3D model. In our experiments, we use the AnyLogic (version 8.7) platform that has high visualization capabilities for both 2D and 3D development. Depending on the complexity and details needed, the development of the environment may take a few hours or a few days. Figure 1 provides examples of a 2D and a 3D view of a hemodialysis unit that were created based on the actual floor plans of the dialysis unit.

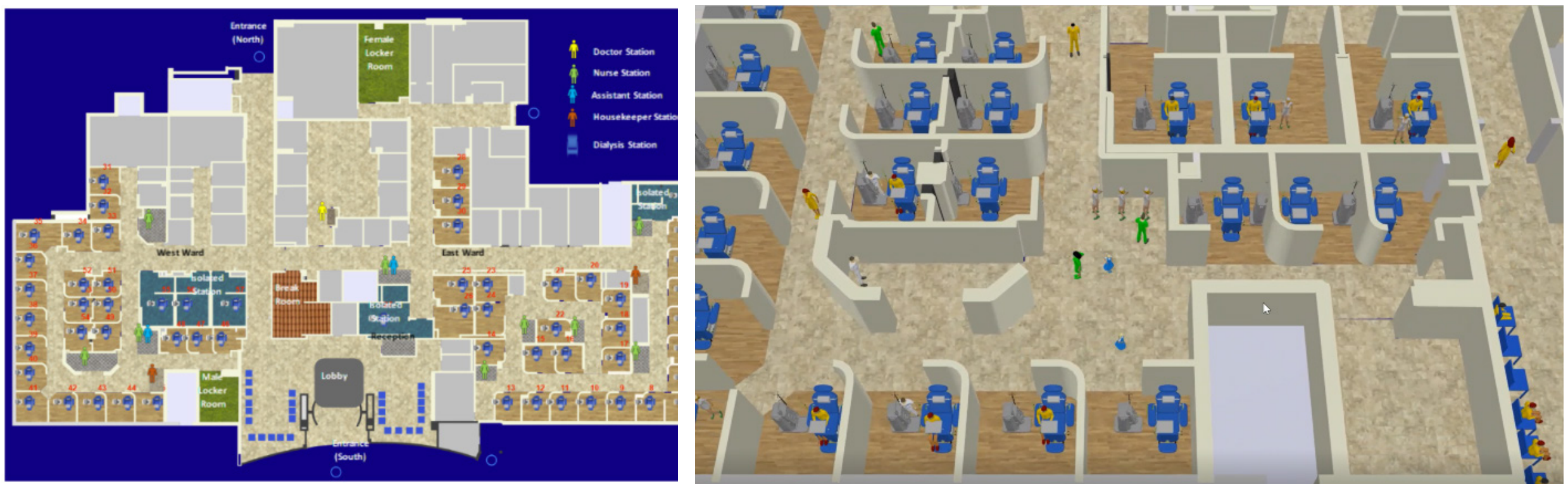

(a) (b)

**Figure 1. 2D/3D simulation environment of a dialysis unit**

## 2.2. Defining Agent's Workflow

In the second step, existing workflow and schedule data of the selected settings are defined to allow modeling of agents' regular and stochastic movements in the environment during a period of time (the study timeframe), e.g., a day or week. Workflow, spatial and temporal data, and schedule information can be collected from managers or staff who work in the selected setting. We use AnyLogic's pedestrian library to model agents' movements according to their predefined daily/weekly workflow and schedules. In this library, pedestrians move according to a social force model. In the social force models, the mass of every individual, desired speed of the individual within the absence of interactions, the direction of movement toward attraction points, the force between individuals or individuals, and obstacles are developed within the conservation of momentum equation. Pedestrians take the shortest route, analyze the current environment to avoid colliding with other objects and make decisions about future movements. These are typically similar to what a pedestrian performs in real conditions. Figure 2 shows examples of pedestrian workflows created for dialysis patients and nurses of the dialysis unit. Nurses and patients are generated at a virtual home and attend the hospital based on their weekly schedules. Patients line up and register at the reception, wait for a vacant ready dialysis unit, and go to the bed. Then they receive their treatment, leave the hospital, and return to the virtual room. Nurses receive their daily schedules when they reach the hospital, then they change their clothes in the locker room. They go to their stations, wait for the assigned patients, and circulate among patients up to the end of their shift. In the defined workflow, the schedules need to indicate key locations and waiting times and let movements between areas be controlled by the social forces. Using a random distribution of time around the average waiting time in a location and assuming free-motion or restricted space in stations, rooms, chairs, beds, and queues will provide a more realistic environment in the simulation.

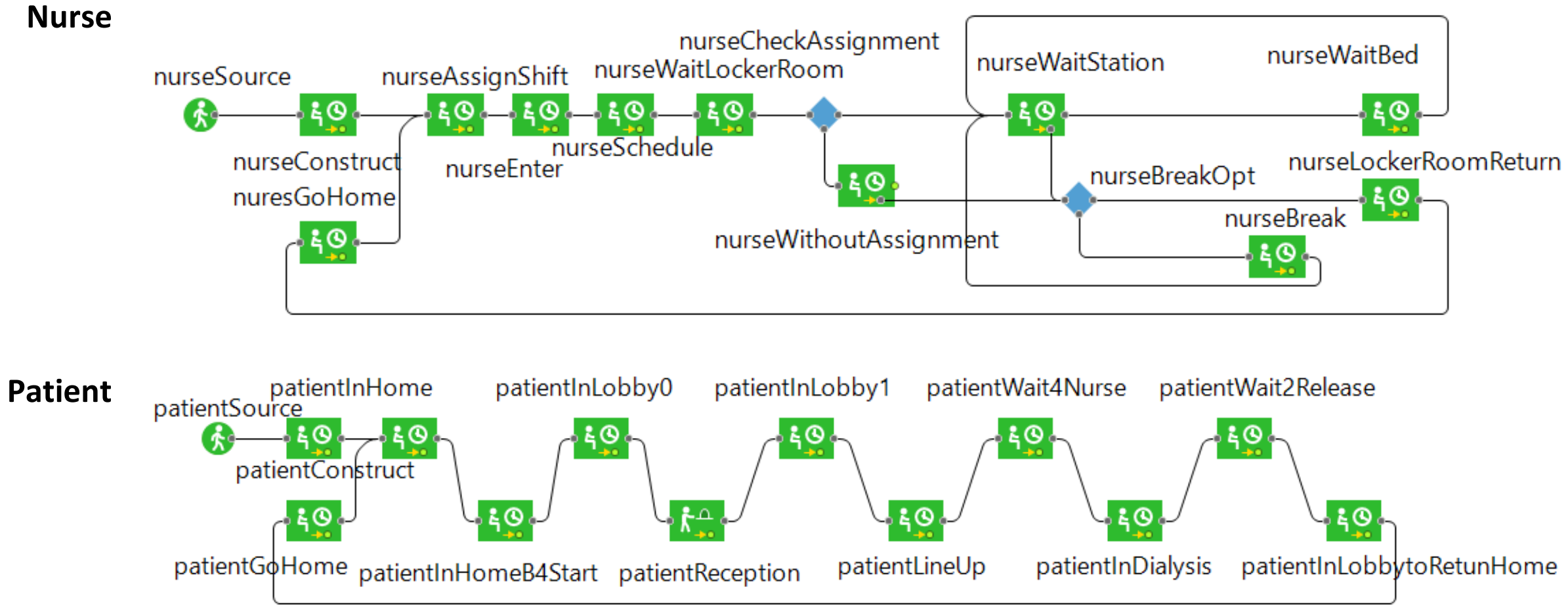

**Figure 2. Workflow of the nurse and patient agents of the hemodialysis unit in the pedestrian model**

Figure 3 shows similar workflows of some of the agents in the Intensive Care Unit (ICU) model. Similar to the nurses in the hemodialysis model, ICU nurses arrive in the unit and prepare for their shift assignments. Then they go to their assigned patient, take the monitoring responsibility from the nurses in the previous shift, and take breaks and lunch while their adjacent colleague watches their assigned patient. ICU physician teams are composed of attending physicians, critical care fellows (CCFs), and residents. They arrive in the ICU and get their teams in a handover room to plan for the day. Then they check the patients one by one during the day. There is also a second round for patient procedures. Because of task complexities of the physician team, we have used a state diagram that helps to define the characteristics of the physician team activities.

**Nurse (ICU)**

**Physician Team (ICU)**

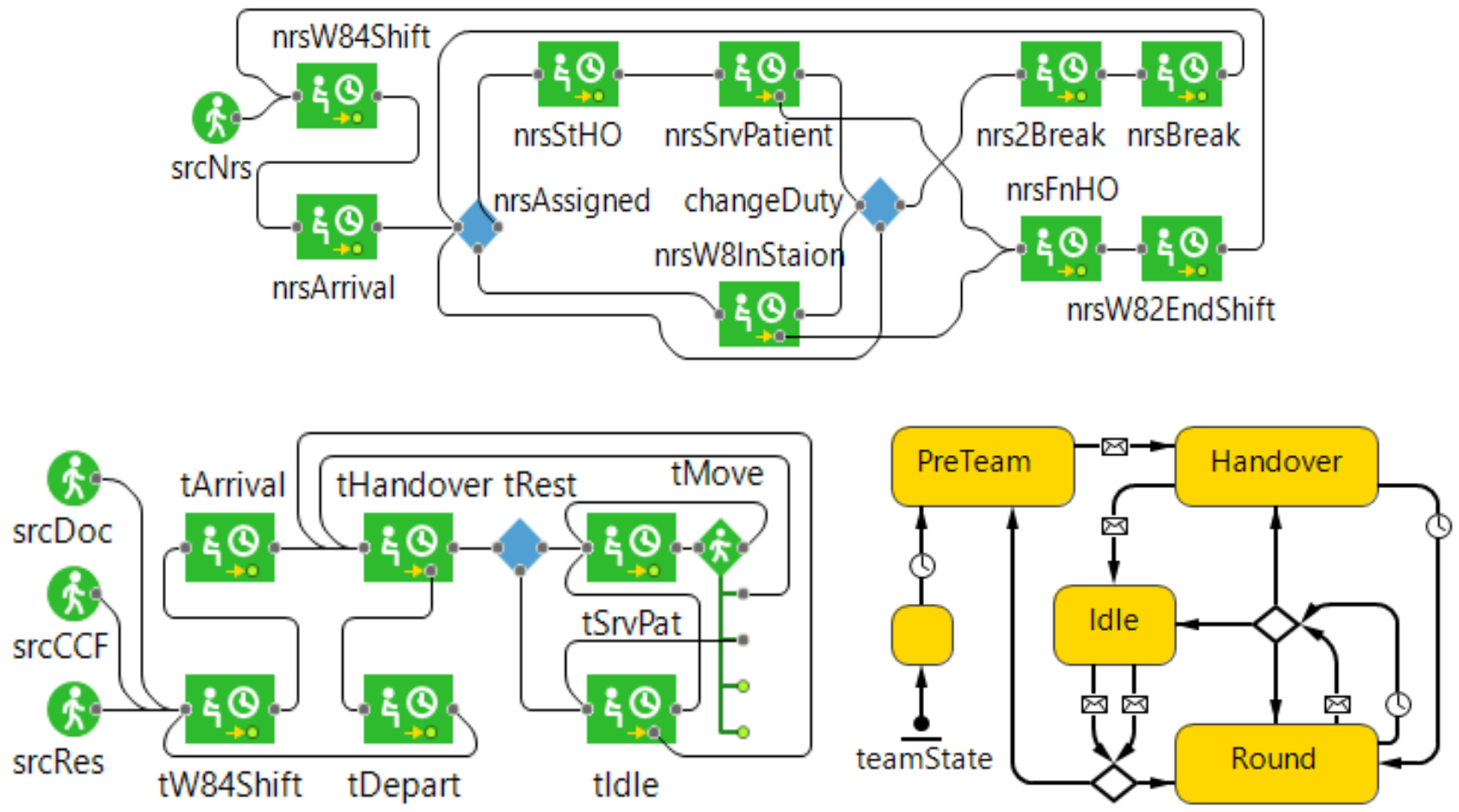

**Figure 3. Workflow of the patient and nurse agents of the intensive care unit (ICU) in the pedestrian model**

The workflows above are the minimum activities of the sample agent types that are presented here, due to their daily schedules or duties. Thus, these workflows are different for each agent type and correspond to the actual schedules of the specific type of healthcare worker in the modeled environment.

Another example is the workflow of mass gathering in the Hajj ritual. In a part of this ritual, pilgrims go to the Masjid-al-Haram (a holy shrine place) and circulate, pray, and walk in specific predefined places. Each pilgrim performs his or her ritual individually in a particular order. The total number of pilgrims in the Masjid can reach up to 60,000. They are in different places in the modeled physical environment. To simulate this workflow, we defined a typical place-duty flowchart for pilgrims that controls the order of the ritual for each individual. Figure 4 shows this flowchart, as well as an abstract view of a part of the physical environment.

**Pilgrim in Hajj ritual**

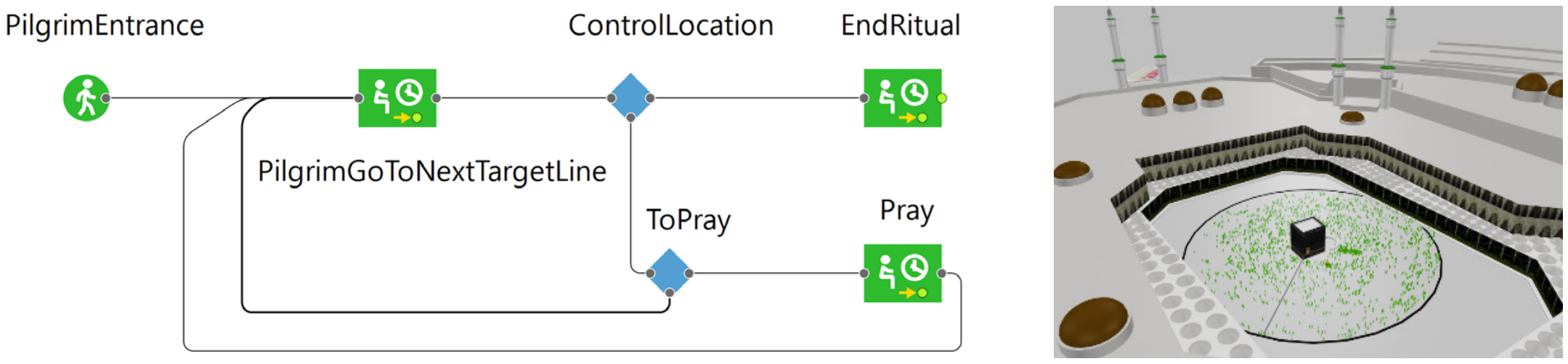

**Figure 4. Workflow of the pilgrims in Hajj ritual in the Masjid-al-haram in the pedestrian model**

## 3. Formulating Contacts and Computing Contacts Mixing Matrices

### 3.1. Conceptual Contacts Logging

Consider two sets of people: set $A$ with a total population of $n_A$ and set $B$ with a total population of $n_B$. Depending on the scope and purpose of the model, the assumption is that these agents live, work, play, gather, interact, etc. together in an environment for a certain period of time (the simulation time frame). Each member of set $A$ can have $n_B$ unique one-to-one contacts with members of set $B$. Similarly, each member of set $B$ can have $n_A$ unique one-to-one contacts with members of set $A$. At the same time, each member of set $A$ can have $n_A - 1$ unique one-to-one contacts with other members of set $A$, and similarly, each member of set $B$ can have $n_B - 1$ one-to-one contacts with every other member of set $B$. Mathematically, the count of all unique contacts that involves one member from each set is $n_A \times n_B$; and the count of unique contacts that involves people from the same set for sets A and B is $\binom{n_A}{2}$ and $\binom{n_B}{2}$ respectively. Thus, the maximum possible unique contacts in this environment that involves only these two agent sets can be expressed as the summation of all three expressions above:

$$n_A \times n_B + \binom{n_A}{2} + \binom{n_B}{2} = n_A \times n_B + \frac{n_A!}{(n_A - 2)! \times 2!} + \frac{n_B!}{(n_B - 2)! \times 2!} \tag{1}$$

Alternatively, it can be assumed that both sets combined together would create a superset in which two members can be in contact with each other, and yield another form of the same formula:

$$\binom{n_A + n_B}{2} = \frac{(n_A + n_B)!}{(n_A + n_B - 2)! \times 2!} \tag{2}$$

In situations such as recurring hospital visits by patients (during which specific nurses serve the patients), coworkers who serve in two adjacent spots in a series of workflows, and coworkers who share an office or work in the same room, the number of contacts between those individuals increases. According to the World Health Organization (WHO), the risk of disease transmission increases in contacts in which the distance between the two individuals is less than a certain threshold. The duration of the contact also directly affects the risk of disease transmission during the contacts. Thus, contact parameters such as the frequency of contacts, the distances between the individuals during the contacts, and the duration of the contacts can determine the risk of disease transmission between people in an environment. It is difficult and time- and resource-consuming to follow every individual and measure the contact parameters in a real environment. This calls for a proper alternative method of logging every unique contact and its corresponding parameters.

In the hemodialysis example above, the contact parameters for every contact per individual are recorded using a data structure, namely *contact log*, that captures contact parameters. Each contact log is a set $C_{ij} = [count_{ij}, dur_{ij}, dist_{ij}]$ that keeps the count of contacts ($count_{ij}$), cumulative duration of the contact ($dur_{ij}$), and the average distance between individuals ($dist_{ij}$) during the contact, where $i$ , $j$ represent the indices of the individuals in sets $A$ and $B$, respectively. Due to the variation of these parameters by time, a timestamp should also be logged with each corresponding contact parameter.

Assuming other parameters (such as workflow, shifts, etc.) are constant, by increasing the modeling timeframe and thus, the span of recording contacts data, the cumulative count and duration of contacts would increase, and the average distance between the agents involved in the contacts would consist of more data points. Knowing these cumulative contact parameters from every agents' point of view, the average (normalized) contact parameters for every set (i.e. every agent type) is calculated, which simply aggregates the agent-level contact parameters into a normalized set of agent-type to agent-type contact parameters. This can be done via the following equations:

$$Count_{A-B} = \frac{\sum_{i=1}^{n_A} \sum_{j=1}^{n_B} count_{ij}}{n_A \times n_B} \qquad (3)$$

$$Count_{A-A} = \frac{\sum_{i=1}^{n_A} \sum_{j=1}^{n_A} count_{ij,i\neq j}}{2 \times \binom{n_A}{2}} \tag{4}$$

$$Duration_{A-B} = \frac{\sum_{i=1}^{n_A} \sum_{j=1}^{n_B} dur_{ij}}{n_A \times n_B} \tag{5}$$

$$Duration_{A-A} = \frac{\sum_{i=1}^{n_A} \sum_{j=1}^{n_A} dur_{ij,i\neq j}}{2 \times \binom{n_A}{2}} \tag{6}$$

$$Average\ Distance_{A-B} = \frac{\sum_{i=1}^{n_A} \sum_{j=1}^{n_B} dist_j \times dur_j}{n_A \times \sum_{i=1}^{n_B} dur_i} \tag{7}$$

$$Average\ Distance_{A-A} = \frac{\sum_{i=1}^{n_A} \sum_{j=1}^{n_B} dist_j \times dur_{j,i\neq j}}{n_A \times \sum_{i=1}^{n_A} dur_i} \tag{8}$$

where $i$ and $j$ represent the index of individuals in sets $A$ and $B$, respectively, $Count_{A-B}$ and $Count_{A-A}$ are the average count of recorded contacts between sets $A$ and $B$, and the same set $A$, respectively, and the $Duration$ represents the cumulative duration of contact. The above equations are symmetric, thus:

$$Count_{A-B} = Count_{B-A} \tag{9}$$

$$Duration_{A-B} = Duration_{B-A} \tag{10}$$

$$Average\ Distance_{A-B} = Average\ Distance_{B-A} \tag{11}$$

Figure 5 shows the process of logging contact data between two individuals ($a$ and $b$) in every time-step. This process should be repeated for all individuals in the study to record all contact parameters. Three main components are used to calculate the contacts: 1) An effective radius ($R_e$), which defines the maximum distance by which the two active agents should be within each other; 2) A time-step (minimum time period) by which the distance of the two persons is checked, to see if they are in contact with each other; and 3) A duration that defines the minimum length of effective contact. Although the third component would filter out some of the potential contacts from what would classify as *effective contacts*, it is still required to store the data at the most granular level (component two: minimum time-period) to apply the third criterion. In other words, it is required to keep track of all contacts to identify the effective ones.

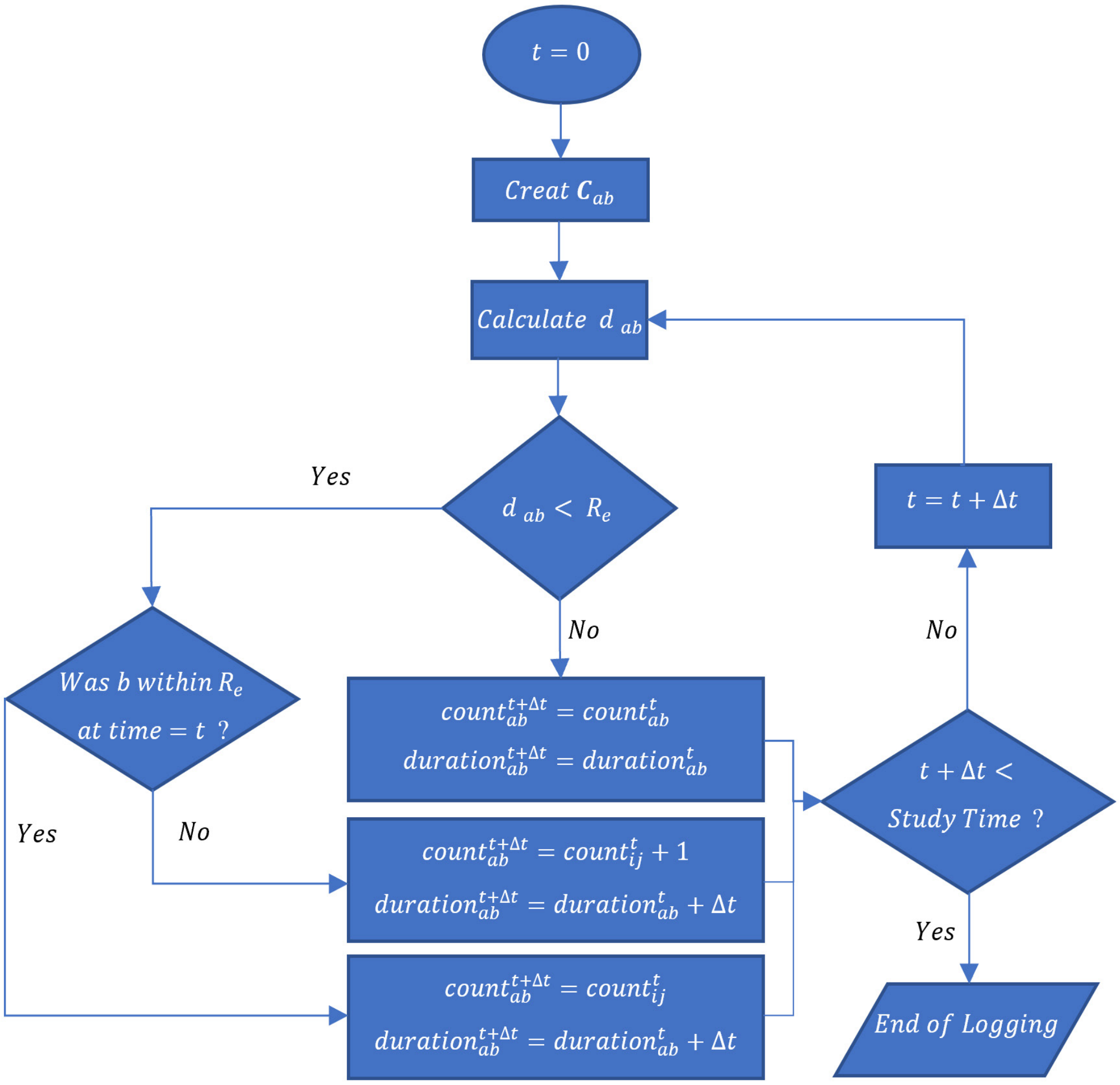

**Figure 5. Flowchart of logging contact parameters between two individuals ($a$ and $b$)**

## 3.2. Detecting and Recording Contact Data

To create a contact matrix, three values should be detect and record: 1) agents that are involved in the contacts; 2) duration of the contacts by calculating the time difference between the start and end times of each contact between pairs of agents; and 3) the average distance between the agents during each contact (between every pair of agents that are in contact with each other). Figure 6 visualizes the contact recording mechanism explained here. Consider an active agent named *host* agent shown in the middle of the circle. Assuming an effective contact radius ($R_e$) of two meters, all agents entering and remaining inside the contact radius for the specified time-step will be recorded. In this figure, members of different sets are shown by different colors: Different *agent types*; e.g. nurses, physicians, clerks, etc.

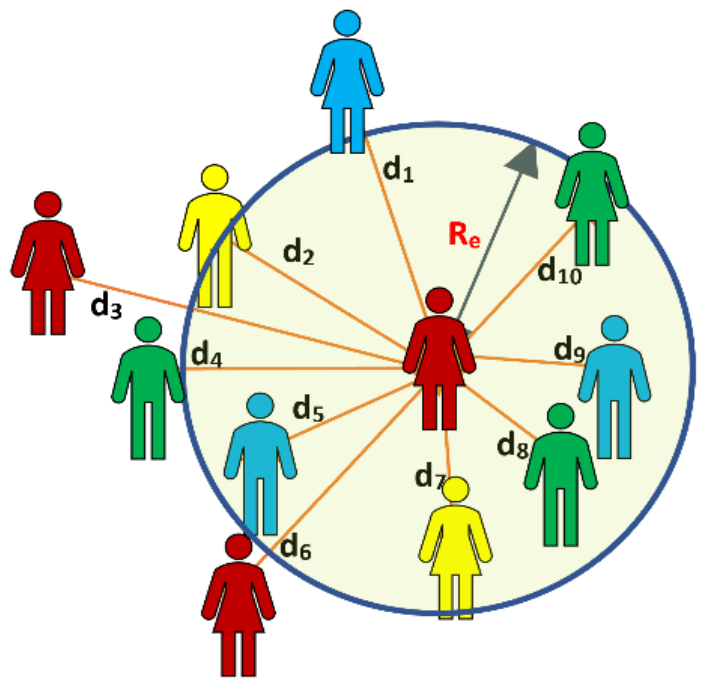

**Figure 6. Contacts within the effective range of an active agent**

Figure 7 shows the physical location of agents in three consecutive time-steps. If an agent is outside of the host's contact radius, it would not be counted as a contact. The *Agents Pair* column shows all the possible contact pairs of all other agents with the host agent, coded in the *Pair Index* column to represent different agent types. The next four columns contain contact information for two consecutive time-steps by which the contacts are being recorded. Note that, because the contact's status at each time-step is checked, the durations of the contacts will be integer numbers (Table 1). If an agent was in contact with the host agent at time *t*, and remained in contact in time *t + 1*, the duration of its contact would be increased by one (see Table 2 and Table 3 below).

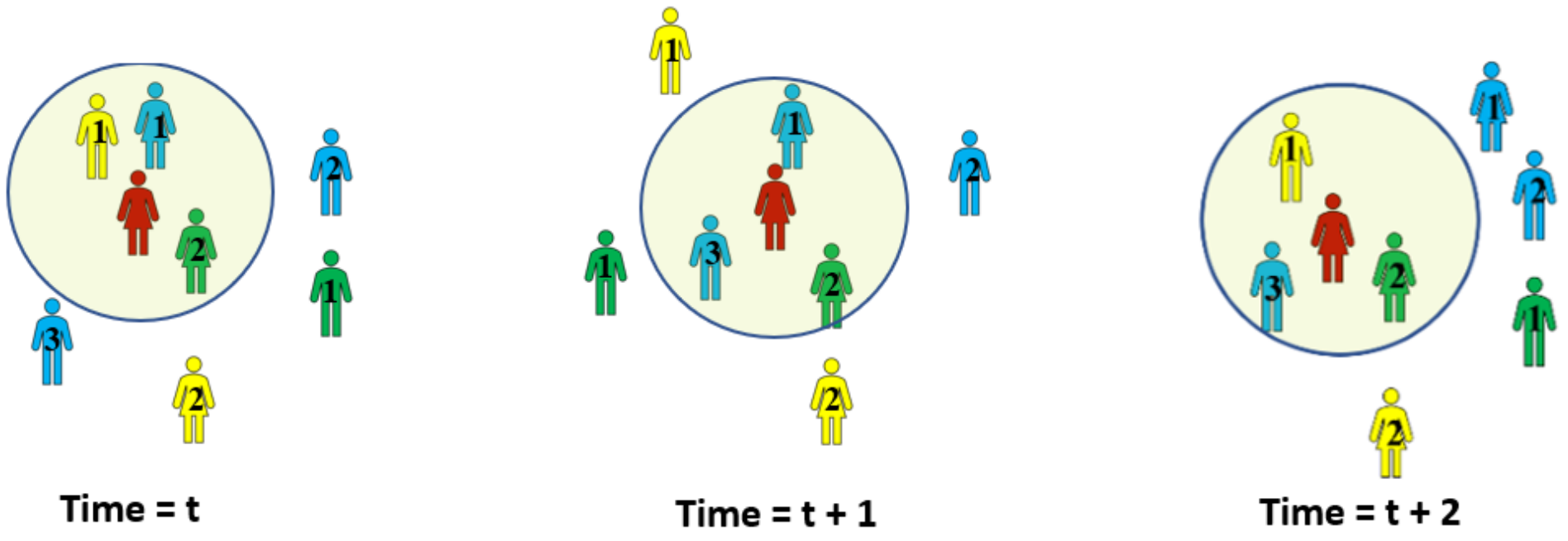

**Figure 7. Agents' physical location according to a host agent's contacts' radius at two successive time-steps in the model**

**Table 1. Contacts parameters at the model time-steps as presented in Figure 6; Time = t**

| Contact ID | Agents Pair | Agent Index | Start Time | Last Updated | Running Duration | Average Distance | Still In- Session? |
|---|---|---|---|---|---|---|---|
| 1 | 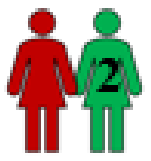 | G2 | t | t | 1 | 1.50 meters | Yes |
| 2 | 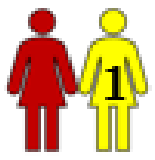 | Y1 | t | t | 1 | 0.90 meter | Yes |
| 3 | 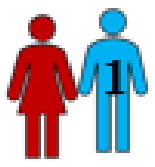 | B1 | t | t | 1 | 1 meter | Yes |

**Table 2. Contacts parameters at the model time-steps as presented in Figure 6; Time = t + 1**

| Contact ID | Agents Pair | Agent Index | Start Time | Last Updated | Running Duration | Average Distance | Still In- Session? |
|---|---|---|---|---|---|---|---|
| 1 | 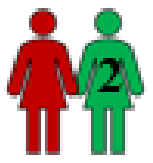 | G2 | t | t + 1 | 2 | 1.75 meters | Yes |
| 2 | 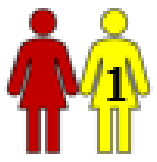 | Y1 | t | t | 1 | 0.90 meter | No |
| 3 | 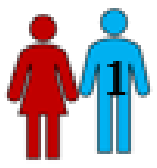 | B1 | t | t + 1 | 2 | 0.90 meter | Yes |
| 4 | 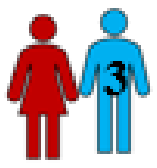 | B3 | t + 1 | t + 1 | 1 | 1.75 meters | Yes |

Now assume that in time *t + 2*, agent G2 remains in the host's contacts' radius, agent Y1 moves into the host's contacts' radius, and agent B1 leaves the host's contacts' radius, and B3 also remains in the host's contacts' radius. In this case, the contact duration for G2 will increment to 3, for B1 it no longer is counted as a contact, and for B3, the same contact will have its duration incremented to 2. For Y1 however, a new contact record will be added to the contacts' array and the duration will become 1. This is because although Y1 was in the radius at times t and t + 2, because it was not in the radius at time t + 1, those two contact events are counted as two different contacts.

**Table 3. Contacts parameters at the model time-steps as presented in Figure 6; Time = t + 2**

| Contact ID | Agents Pair | Agent Index | Start Time | Last Updated | Running Duration | Average Distance | Still In- Session? |
|---|---|---|---|---|---|---|---|
| 1 | 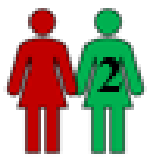 | G2 | t | t + 2 | 3 | 1.70 meters | Yes |
| 2 | 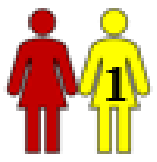 | Y1 | t + 2 | t | 1 | 0.90 meter | No |
| 3 | 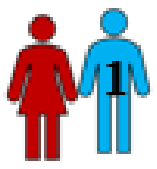 | B1 | t | t + 1 | 2 | 0.90 meter | No |
| 4 | 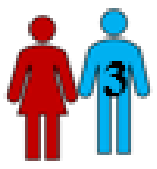 | B3 | t + 1 | t + 2 | 2 | 1.80 meters | Yes |
| 5 | 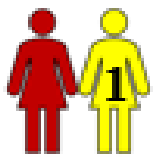 | Y1 | t + 2 | t + 2 | 1 | 1.10 meter | Yes |

By recording contacts data at this granular level, building the contacts matrix at any desired level of abstraction is applicable, and different inclusion criteria to create the contacts matrix (e.g., to include only the contacts that are in-progress and longer than a specific time period) is possible. This data structure is basically a four-dimensional array that records all details (dim 4) of every single contact (dim 3) for every contacted pair of agents (dim 2) for all existing agent-types that are used in computing the contacts matrix for (dim 1). Dimensions 1 and 2 are similar because both are agents of different agent-types that are cross-contacting with each other.

The used data structure logs every contact between any two pairs of agents at each time-step. Contacts can continue in consequent time-steps until the first time-step in which the two corresponding agents are no longer in the proximity of each other. These continued contacts will only have an updated finish timestamp until they are in-session. Once the contact is no longer in-session, the finish timestamp will no longer update, and the contact is considered a past contact. Once the same agents are in contact again, a new contact record will be generated and logged for the same pair of agents with a new start timestamp. The same process regarding the finish timestamp continues until the contact between the two agents is no longer in-session again. This data structure is capable of quickly reporting the count of risky contacts, i.e., the ones that last more than a certain amount of time. Contacts that are in-session for less than that duration are still risky – but maybe less so.

## 4. Running the Simulation and Extracting the Contacts Matrix

During the simulation runs, the contacts data is granularized at the level of each contact. The same two agents can have multiple contacts during the simulation time and each of those contacts will have a separate record in the contacts' array. Furthermore, this data is captured at both ends of the contact for both agents involved. These contacts data are recorded for each active agent in the model. To be able to use this data, it is required to aggregate and summarize them at a more abstract level. By summarizing the data into more aggregated and normalized averages, contact matrices can be obtained.

### 4.1. Aggregating Contacts at Individual Agent Level

To aggregate the contacts data at the individual agent level, the information recorded at each agent's contacts array is summarized. This is done by compressing the four-dimensional array of contact details at its third dimension. For that, all the pairs of contacted agents with a host are formed, and the contact information for each agent pair is derived from the host's contact array. For the example illustrated in Figures 6 and 7 (assuming that the simulation time is only three time-steps), this list would contain Host-G2, Host-Y1, Host-B1, and Host B3. This is because G1, Y2, and B2 were never in the radius of the contact of the host agent. Now for the contacted agent pairs listed, all the contacts are summed up and normalized from the host's contacts' array: 1) The count of contacts between the host agent and each other contacted agents, is the number of recorded contacts that involved the contacted agent in the array (which is the averaging denominator); 2) The cumulative duration of contacts between the host and every other contacted agents, that is the sum of all contact durations between the two agents; and 3) The average distance in the contact distances between the host agent and each of the other agents, which can be obtained through a weighted average formula based on the duration of each contact. Thus, the sum-product of the average distances per contact and the corresponding contact duration, divided by the cumulative durations of the contacts with the same agent. For the example illustrated in Figure 6 and Table 1, this would yield a contacts Table as depicted in Table 2.

**Table 4. Host agent's aggregated contacts at the agent level**

| Contacted Agent | Count of Contacts | Cumulative Duration of Contacts | Average Distance of Contacts |
|---|---|---|---|
| G2 | 1 | 3 seconds (100%) | 1.70 meters |
| Y1 | 2 | 2 seconds (67%) | 1 meter |
| B1 | 1 | 2 seconds (67%) | 0.90 meters |
| B3 | 1 | 2 seconds (67%) | 1.80 meters |

This data can be generated for each of the active agents. From there, it's possible to generate a large $m \times m$ matrix in which all agents are cross-listed (all agents are listed on rows and columns), where $m$ is the total number of agents in the system that had been involved in any type of such recorded contact (sum of all $n_x$ that were involved). For each of the contacts data (count, average duration, and average distance), we would need a separate matrix, with the $m \times m$ rows and columns (the active agents) remaining constant. For example, the average duration matrix for the examples illustrated in Figure 7 and Table 1 is presented in Table 5. Please note that the contacts between agents that do not involve the host were not discussed earlier in either of the tables but are calculated for the matrix presented in Table 5, which is calculated from the situations presented in Figure 7.

**Table 5. Contacts matrix for the total duration of the contacts between active agents– Individual agents' level**

| | H | G1 | G2 | Y1 | Y2 | B1 | B2 | B3 |
|---|---|---|---|---|---|---|---|---|
| **H** | – | 0 | 3 | 2 | 0 | 2 | 0 | 2 |
| **G1** | 0 | – | 0 | 0 | 0 | 0 | 2 | 1 |
| **G2** | 3 | 0 | – | 2 | 1 | 2 | 0 | 2 |
| **Y1** | 2 | 0 | 2 | – | 0 | 1 | 0 | 1 |
| **Y2** | 0 | 0 | 1 | 0 | – | 0 | 0 | 0 |
| **B1** | 2 | 0 | 2 | 1 | 0 | – | 1 | 0 |
| **B2** | 0 | 2 | 0 | 0 | 0 | 1 | – | 0 |
| **B3** | 2 | 1 | 2 | 1 | 0 | 0 | 0 | – |

Forming this matrix requires the contacts data recorded for all other agents in the environment (not just for the host and not just for the small sample of agents that appear in Figure 6). As can be seen in Table 5, the diagonal of the matrix (which represents an agent contacting itself) is undefined. Furthermore, since each contact is captured at both its ends by the two agents involved in the contact, this matrix is symmetric. Although, from a numeric standpoint, the mirrored numbers might not exactly match due to time-step approximation and other computational limitations in the simulation platform.

## 4.2. Aggregating Contacts at Agent Type Level

The process of aggregating contacts data is followed by further compressing the data array on its second dimension. This would yield a more abstract construct (i.e., contacts matrix) at the individual agents' level (host agents), which contains aggregate numbers describing contact properties (i.e., average cumulative duration in contact, and average distance) per agent type. The average count of contacts would be the total number of contacts of the host agent, with all agents for a specific period (e.g., for the whole simulation period). The average cumulative duration of contacts is the average of all cumulative durations of the contacts, based on the number of agents contacted by each agent type. And the average distance in the contacts is a weighted average of all average contact distances, adjusted for the cumulative contact duration. For the example presented in Figure 7 and Table 1, the more abstract data table is presented in Table 6.

**Table 6. Host agent's aggregated contacts at the agent level**

| Contacted Agent Type | Normalized Count of Contacts | Average Cumulative Duration of Contacts | Average Distance of Contacts |
|---|---|---|---|
| Green | 0.5 | 1.5 seconds (50%) | 1.75 meters |
| Yellow | 1 | 1 second (33%) | 1 meter |
| Blue | 0.67 | 1.33 seconds (44%) | 1.35 meters |

(Once again, please note that the count of contacts is presented just to help with digesting the algorithm and will not be used further in the aggregations.)

This data can be generated for each of the active agents that their contact details have been recorded. The matrix for this level of aggregation will have one of its dimensions aggregated to the agent types, instead of individual agents. Thus, here, an *m x n* matrix, in which m is the count of total individual agents with recorded contact data, and *n* is the count of agent-types that have been included (Table 7) is obtained.

**Table 7. Contact matrices for the duration of the contacts between active agents that have been in contact – Individual agents' level aggregated by agent type**

| | H | G1 | G2 | Y1 | Y2 | B1 | B2 | B3 |
|---|---|---|---|---|---|---|---|---|
| **H** | – | 0 | 3 | 2 | 0 | 2 | 0 | 2 |
| **G** | 1.5 | 0 | 0 | 1 | 0.5 | 1 | 1 | 1.5 |
| **Y** | 1 | 0 | 1.5 | 0 | 0 | 0.5 | 0 | 0.5 |
| **B** | 1.33 | 1 | 1.33 | 0.67 | 0 | 0.33 | 0.33 | 0 |

This matrix is no longer symmetric because it is aggregated on one of its dimensions (rows or columns) while the other dimension is kept at the previous granular level. Moreover, the diagonal will no longer be undefined, because each agent can have contacts with other agents of its own type.

## 4.3. Aggregating Contacts for all Agents

A higher level of abstraction can be achieved by aggregating contacts data by further compressing the data array on its first dimension. To do this the contacts data for all agents from each agent type are aggregated and contacts tables for different pairs of agent-types are computed. This approach would eventually lead to the contacts matrix that depicts the average contact, duration, and distance between each of the agent-types with other agent-types.

Applying these steps to the example presented in Figure 7 and Table 1 would yield an $n \times n$ matrix represented in Table 8, where n is the count of agent types for active agents that have ever been in contact with another active agent.

**Table 8. The contacts matrix for the duration of the contacts between active agents that have been in contact – Agents' type level**

| | **H** | **G** | **Y** | **B** |
|---|---|---|---|---|
| **H** | – | 1.5 | 1 | 1.33 |
| **G** | 1.5 | 0 | 0.75 | 1.17 |
| **Y** | 1 | 0.75 | 0 | 0.33 |
| **B** | 1.33 | 1.17 | 0.33 | 0.34 |

As can be seen in Table 8, the diagonal is no longer undefined. This is because it represents an agent's average number of contacts with other agents in the same set (i.e., of the same agent type), unless one of the sets has only one agent (e.g., H representing the host agent in the previous example in Figure 7). This matrix is symmetric because each contact is recorded at both ends by both involved agents. Then, as these data are summarized in two dimensions, the symmetric nature of the data is revealed again.

Table 9 shows the extracted contacts matrix for the dialysis unit that was introduced earlier.

**Table 9. Average count* (normalized frequency) of contacts per day in the hemodialysis simulation model**

| | Patient | Clerk | Housekeeper | Assistant | Nurse | Physician | Nephrologist |
|---|---|---|---|---|---|---|---|
| **Patient** | 0.12 | 0.02 | 0.70 | 0.12 | 0.35 | 0.00 | 0.00 |
| **Clerk** | 0.02 | 0.70 | 1.75 | 2.22 | 2.10 | 0.00 | 0.00 |
| **Housekeeper** | 0.70 | 1.75 | 2.8 | 13.8 | 7.6 | 0.00 | 0.00 |
| **Assistant** | 0.12 | 2.22 | 13.8 | 77.1 | 41.1 | 0.12 | 0.12 |
| **Nurse** | 0.35 | 7.6 | 7.6 | 41.1 | 15.4 | 0.05 | 0.05 |
| **Physician** | 0.00 | 0.00 | 0.00 | 0.12 | 0.05 | 0.00 | 0.35 |
| **Nephrologist** | 0.00 | 0.00 | 0.00 | 0.12 | 0.05 | 0.35 | 0.00 |

*** Note: The above contact matrix is for 308 patients, 6 clerks, 10 housekeepers, 30 assistants, 110 nurses, 10 physicians, and 12 nephrologists.**

The contacts matrix above contains the more aggregated data after being compressed into two dimensions. It shows agent-type to agent-type contact rates per day in the dialysis unit. As explained earlier, this matrix is symmetric because the diagonal shows the average contact rate between an agent with all other agents from the same type. The extracted values in this hospital are the same order of magnitude as the values of the measured contact in another hospital (Baek et al., 2020). Using this approach enables us to produce the contact parameters in different time scales. For example, Figure 8 demonstrates the time series of the length of contacts during a day of work in the dialysis department. These values, which are usually presented as a single average daily, are time-dependent and vary during a period of time. Having this granularity, more detailed epidemic models can be developed to investigate disease transmission and define more effective policies to reduce the spread of diseases.

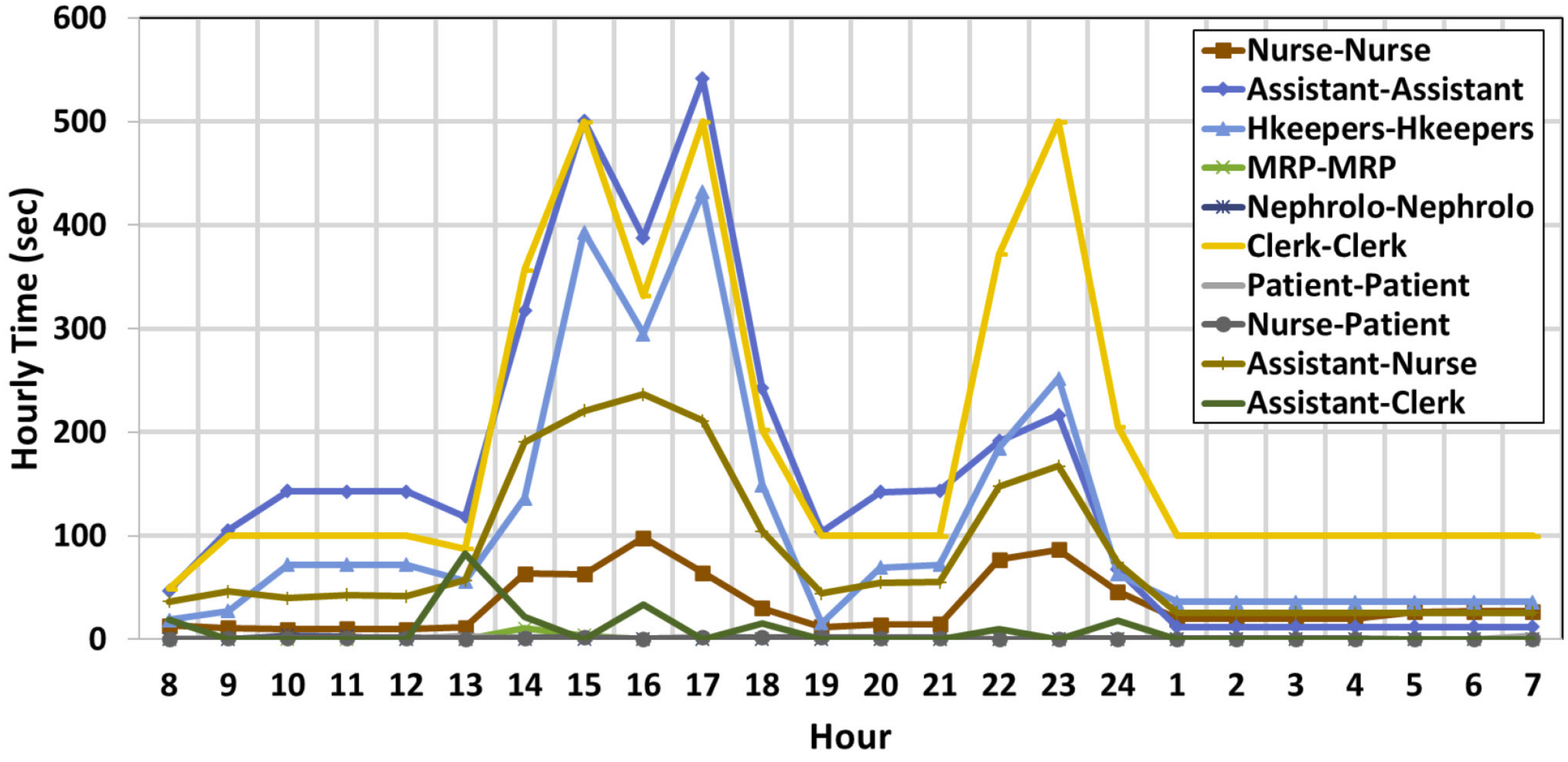

**Figure 8. Time series of hourly cumulative time of connections between agents in the dialysis department**

An alternative way of looking into logging the contact parameters would be to quantify the lengths of contacts into chunks of effective contacts. This can be done using a conditional probability (Bayesian method) to calculate the compound probability of viral transmission in a contact, given the probability of viral transmission in a single effective contact, and the number of effective contacts (lengths of contacts divided by the duration of a single effective contact). This process can be repeated for each pair of agents, using the following equation:

$$Probability\ of\ virus\ transmission = 1 - (1-p)^f \qquad (12)$$

where $p$ is the base probability of virus transmission in one effective contact, and $f$ is the quantified frequency of the effective contacts that were calculated based on the total length of all effective contacts of the corresponding agent-types.

**Table 10. Average chunks of effective contacts per day in the ICU simulation model**

| | AH Staff | CCF | Doctor | Nurse | Resident | RRT |
|---|---|---|---|---|---|---|
| AH Staff | 5.14 | 0.08 | 0.20 | 0.22 | 0.15 | 0.03 |
| CCF | 0.08 | 5.76 | 33.76 | 0.44 | 25.28 | 0.44 |
| Doctor | 0.20 | 33.76 | 3.49 | 0.74 | 47.51 | 1.39 |
| Nurse | 0.22 | 0.44 | 0.74 | 0.83 | 0.63 | 0.00 |
| Resident | 0.15 | 25.28 | 47.51 | 0.63 | 13.86 | 0.99 |
| RRT | 0.03 | 0.44 | 1.39 | 0.00 | 0.99 | 8.17 |
| Patient | 1.81 | 0.66 | 1.33 | 0.10 | 1.05 | 0.95 |

The matrix presented in Table 10 provides the frequencies that can be plugged into the compound probabilities formula above. This will yield the outcome, which is the probability of viral transmission between each pair of agents based on their contact rates.

The results presented in Tables 8 and 10 were within the expected ranges, according to subject matter experts, in the hemodialysis ward and intensive care unit under this study.

There are more ways in which the logged data can be used. For example, one might want to include the average distance of two agents during a contact. Since the average distance is being recorded for each contact, this is possible by using the same data structure.

## 5. Discussion

In this paper, we proposed a simulation-based method for generating a contacts matrix for disease transmission modeling. We have implemented this methodology in a few settings, including the examination of COVID-19 outbreaks in a dialysis setting (Tofighi et al., 2021). The approach is flexible enough to be used in all settings where people are interacting together. The size of space and time do not restrict the performance of the approach. The main requirements of this method are the predefined schedules and space of motions. Movements and interactions are indicated based on the physical conditions that, in the real world, control the dynamics. Further studies on the theory and application of this method are needed to better understand its capabilities and limitations. A number of items related to this method are briefly discussed below.

1. The numbers in the calculated contacts mixing matrix have to be reflective of the risk and frequency by which agents contact each other to transmit the disease at different levels of granularity (agent to agent, agent to agent type, and agent type to agent type). These matrices can focus on different contacts parameters based on the corresponding application, such as: 1) Count of effective contacts (effective, by some predefined characteristics by the user; e.g., longer than a certain length of time, and average distance below a certain level); 2) The normalized cumulative length of contacts (durations); 3) The normalized average distance of the agents while being in contact; and 4) A combination of all contact information above that represents a *risk factor frequency* from this data. That can help viral transmission estimates in longer-term simulation models – which can be yielded through different equations in different disciplines. For example, in some sources, time / distance or time / distance squared is a measure for such.
2. Some very lengthy contacts are much riskier than shorter contacts. For example, a contact that takes place for over an hour should be treated as much riskier than a contact that lasts for only 15 minutes, although the count measure for both is only one contact. Likewise, contacts in which the average distance of the agents involved is shorter, are riskier than the contacts in which agents are farther away from each other.
3. In models in which there is a separate disease transmission module that takes care of simulating virus transmission and propagation, having the contacts data recorded at this level of granularity can be helpful to make the implementation of the disease transmission module more straightforward. The contacts matrix granular data can work hand-in-hand with a viral transmission module that needs to identify Effective Contacts. These are contacts that have lasted for over a certain amount of time – e.g., five minutes according to the CDC (2020) – in which the average distance between the two active agents has been below a given effective radius (e.g., two meters). Relying on the contacts matrix data that has already been collected in the model saves a great deal of additional coding that would have been needed otherwise for a viral transmission module to keep track of active agents' contacts separately.

4. Contact matrices can help with fast simulation modeling because they can be fed into more abstract models that do not involve agents' workflow and movement details. This makes them capable of simulating much longer time frames in a shorter period of time.

5. The aggregated data at either level of abstraction as mentioned in the previous subsection has a 2-dimensional array (matrix) structure for each of the contact elements (i.e., the count of contacts, duration, and average distance). This matrix is sometimes called the adjacency matrix in graph theory, the proximity matrix, or the Who Acquired Infection from Who (WAIFW) matrix in epidemiology. A visual representation can better explain how the big picture looks in terms of the risk of contacts, and where the decision-makers should focus their attention for the most effective results.

6. Transmission of diseases depends on many parameters such as route of transmission (e.g., air droplets, vector-born, blood, unclean wound), the transmission cycle, and reservoir. For different types of diseases, the route of transmission can be related to the contacts between either the infected individuals or infected environments (e.g., un-sanitized surface or infected air) and exposure to the susceptible individuals. Based on disease parameters such as count of contact, duration of the contact, and average distance of the agents during the contact, are important in disease transmission. For example, hospital endemic infections such as Methicillin-resistant Staphylococcus aureus (MRSA) can be transmitted by direct skin-to-skin contact or contact with shared items or surfaces (CDC, 2015). And for COVID-19, it is stated that someone who is within 6 feet of an infected person for a cumulative total of 15 minutes or more over 24 hours starting from 2 days before the illness onset (or, for asymptomatic patients, 2 days before testing specimen collection) is prone to infection (CDC, 2020). The method presented in this paper provides detailed information about contacts at a granular level as much as precision is required for studying the mechanism of disease transmission. By using this methodology, the mathematical models of disease transmission would provide more realistic transmission parameters that can be related to the counts of contacts, duration of contacts, average contact distance, or any compositions of these contact parameters. By defining locations, surfaces, or any parts of an environment as an active agent in the model, the user can calculate the parameters of the contacts with any arbitrary criteria.

7. With the recent advancements in artificial intelligence and machine learning methods, computer simulations can take advantage of an artificial intelligence component that can help to find optimized solutions for a given problem in several problem domains, including public health. For example, Asgary et al. (2020) have applied this method to their drive-through vaccination clinic simulation model to find optimized layouts and staffing levels for given clinic settings and layouts.

8. The contact matrices generated in this study can be created, not only for human-to-human contacts, but also for human-to-object contacts. Therefore, for disease models where contacts between humans and objects are of importance, this methodology can be used to estimate contact patterns.

## 6. Conclusion

In this paper, we introduced a simulation-based approach to generating contact matrices that can be used in disease modeling. The generated matrices can be used to model disease transmission in special settings or generating contact information for macro-scale disease models when contact data about workplaces, community settings, schools, or places of worship, or sport facilities are needed. This approach is particularly useful for situations when access to people or places is not possible, either because of the risk or because of their temporary closures.

Furthermore, once the simulation has been developed, it is possible to generate many varieties of contact matrices under different disease prevention measures in the settings.

A combination of this method and advanced artificial intelligence and machine learning approaches can be used for designing workflows, or environment settings, with less risk of disease transmission. Reinforcement Learning can be employed as a simulation-based method to minimize the number of contacts by optimizing the schedule of agents and the location of accessories, given the environmental constraints such as various geographical scales or layouts. A recent application of agent-based reinforcement learning for achieving optimization of environment setting is presented in the simple objective of hide-and-seek (Baker et al., 2020). This can be modified to optimize the goal of the minimum risk of contacts as the payoff function, that allows the underlying model to calculate and suggest a combination of policy choices and strategies that would help to achieve better results in terms of disease transmission and pandemic control. For example, there are some spatial and temporal restrictions such as the time of shifts or treatment, and some flexibility in the time of attending to the department, selecting the weekly schedule, and choosing the station in the dialysis department. By defining the treatment with fewer contacts in a week as a reward function, it is possible to reach the optimum weekly schedule for a safer operation of the department. Using Reinforcement Learning in the Agent-Based Simulation model will enable the decision-makers to find efficient policies and define the necessary rules for the management of the department. This hybrid simulation has had the capability in the queue network systems (Fuller et. al, 2020).

More applications of this method can generate contact matrices for different settings that can help to define a standard benchmark for comparing infectious disease risk levels in different environments or even similar environments in different locations. As more simulation and contact matrices based on simulations are created, better generalizations can be made across different settings.